# Musical Instrument Recognition by XGBoost Combining Feature Fusion

Yijie Liu [1], Yanfang Yin [1], Qigang Zhu [1,*] and Wenzhuo Cui [1]

1   Shandong Univ Sci & Technol, Dept Elect Engn & Informat Technol, Jinan 250031, Peoples R China; yijie-liu@sdust.edu.cn(Y.L.); yyfang@sdust.edu.cn(Y.Y.); 2445038157@qq.com(W.C.)

*   Correspondence: skd992356@sdust.edu.cn

**Abstract:** Musical instrument classification is one of the focuses of Music Information Retrieval (MIR). In order to solve the problem of poor performance of current musical instrument classification models, we propose a musical instrument classification algorithm based on multi-channel feature fusion and XGBoost. Based on audio feature extraction and fusion of the dataset, the features are input into the XGBoost model for training; secondly, we verified the superior performance of the algorithm in the musical instrument classification task by com-paring different feature combinations and several classical machine learning models such as Naive Bayes. The algorithm achieves an accuracy of 97.65% on the Medley-solos-DB dataset, outperforming existing models. The experiments provide a reference for feature selection in feature engineering for musical instrument classification.

# 1. Introduction

The musical instrument classification task requires the classification system to be able to recognize the musical instrument based on the audio information. This task can be used for tagging music files, emotion recognition and similar music recommendation, which is an important part of the audio signal processing.

In industrial circles, current mainstream music applications and audio search engines can already implement music information retrieval functions, but the implementation method still relies essentially on the tagging information of audio files for retrieval, while information that reflecting the characteristics of music itself, such as timbre, melody and pitch, can be greatly lost[1,2]. And artificial tags often lack the "instrument" attribute, which means that users have difficulty finding accurate results when searching with instrument names as keywords.

In order to mine information from audio signals, it is necessary to match appropriate feature engineering from timbre perception[3]. While features of time domain, frequency domain and cepstral domains are closely related to timbre perception[4-7]. Sikora proposed MPEG-7[8,9], a standardized framework for extracting audio features and descriptions, including audio features such as spectral centroid, Spectral distribution and spectral flatness. Brown et al. obtained good classification results on a dataset of woodwind instruments based on the autocorrelation coefficients of audio[10]. Mar-tin used the excitation and resonance architectures of musical instruments as a re-search point to propose a set of acoustic features that can effectively represent the timbre of musical instruments[11].

In this paper, through literature research, 18 timbre-related features in the time, frequency and cepstral domain are extracted and multi-channel features fusion, and the XGBoost classifier is used to build a more superior musical instrument classification model. And we verify the superior performance of the model by comparing the Classification result with existing research results, mainstream machine learning models and mainstream feature selection strategies.

The main contents of each chapter are as follows:

1. In Chapter 2, we present the characteristics of each feature we select and the ex-traction process. The timbre-related features are explained separately in three parts: time domain, frequency domain and cepstral domain. Then we introduce the multi-channel feature fusion algorithm.
2. Chapter 3 introduces the principle of XGBoost algorithm and its derivation process.
3. Chapter 4 details the implementation steps of the musical instrument classification algorithm, including the audio preprocessing method, the hyperparameter of the model, and verifies the effectiveness of the algorithm by multiple sets of feature comparison experiments and model comparison experiments on the dataset.

# 2. Audio Feature Selection

Currently in the field of musical instrument classification, the strategy of feature selection for feature engineering is mostly carried out in the same domain. This approach has some merits, but the articulatory mechanisms of musical instruments are different, and the single-scale feature learning approach can cause certain limitations[12]. For example, orchestral instruments have obvious resonance structures, but percussion instruments do not have obvious resonance peaks. Therefore, this paper proposes a feature fusion-based instrument feature learning method, which extracts 18 timbral features from the time domain, frequency and cepstral domain, and each feature is evaluated for its mean and variance to obtain 36-dimensional features.

*2.1. Frequency Domain Feature Extraction*

The frequency domain features are vectors generated after some changes to a frame of sound signal, and the spectrum of different musical instruments varies greatly. In this paper, the frequency domain features are extracted from the spectrum of Short-time Fourier Transform (STFT).

For a discrete instrument audio signal $x(t)$ with number of points $N$, we define a short window function $\omega(n)$ shifted on the time axis, and its STFT spectrum is given by:

$$X(m,f) = \sum_{n=0}^{N-1} x(n)\omega(n-m)e^{-j2\pi fn}. \#(1)$$

The Hamming window function is used for $\omega(n)$ in this paper.

As shown in Figure 1, the four plots are the STFT spectrum of flute, piano, tenor saxophone and violin from left to right. We can see that the spectrum of the flute attenuates faster; the violin is a stringed instrument with many harmonics, the material and thickness of the four strings are different, and their respective tones are different; the body of the saxophone makes its timbre mellower than that of the flute. In addition, except for the violin, the number of harmonics of the other three instruments decreases as the pitch increases.

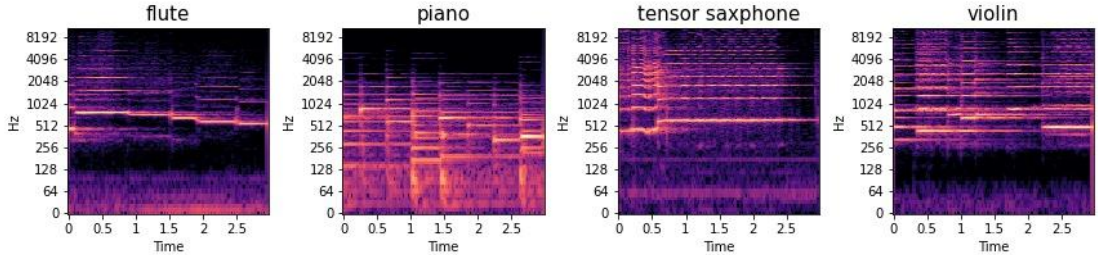

**Figure 1.** Spectrum of each instrument.[1]

2.1.1. Extraction of Spectral Centroid (SC)

The spectral centroid is an important parameter for describing timbre, which refers to the frequency that is weighted and averaged by the energy in a certain frequency range[13]. In the timbral representation of musical instruments, the spectral centroid describes the brightness of the timbre. If the spectral centroid has more low-frequency parts, the timbre of the audio tends to be low. If the high frequency part of the spectral centroid is more, the timbre of the audio tends to be cheerful[14].

Figure 2 shows the images of the spectral centroids of the flute and trumpet as an example. The trumpet has a lower timbre compared to the flute, and its spectral centroid is significantly lower than that of the flute.

---

[1] The code of drawing Figure 1-4, 6 is available at the URL:
https://github.com/Jay-Codeman/Musical-Instrument-Recognition-by-XGBoost.

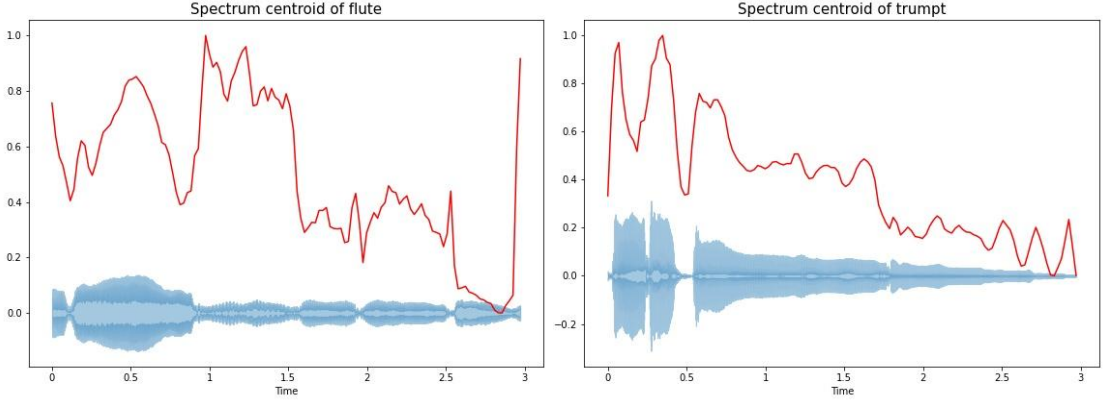

Figure 2. Spectral centroids for flute and trumpet.

First, define the normalized STFT spectrum as:

$$\widetilde{X}(m) = \frac{|X(k)|}{\sum_{k \in K_+} |X(k)|}, \#(2)$$

where $k$ is the frequency index, and the set $K_+$ contains only non-negative frequency index. The spectral centroid is defined as:

$$C_f = \sum_{k \in K_+} k \widetilde{X}(k). \#(3)$$

2.1.2. Extraction of Spectral Spread (SS)

The spread is the difference between the high and low frequencies in a continuous frequency band, and is the brightness of the spectrum mean spread relative to Centroid[15], derived from the spectral centroid $C_f$, and is directly related to the timbre of the audio[16]. The calculation formula of SS is as follows:

$$S_f^2 = \sum_{k \in K_+} (k - C_f)^2 \widetilde{X}(k). \#(4)$$

Generally speaking, the more tones, the louder the sound, and the higher the SS; the lower the pitch, the lower the SS[17].

2.1.3. Extraction of spectral roll-off

Spectral roll-off refers to the critical frequency corresponding to the drop in amplitude to a specific percentage of the total energy of the spectrum[18] (the percentage is set at 85% in this paper) and is a measure of the waveform plot used to distinguish between turbid and clear tones of audio, where most of the energy of clear tones is contained in the high frequency range. Indicates the asymmetry of frequencies in a frame, reflecting the distribution of signal energy over frequencies[19].

$$R_r = arg\left(\sum_{k=1}^{R_r} X_r(k) = 0.85 \cdot \sum_{k=1}^{K} X_r(k)\right), \#(5)$$

where $k$ is the frequency index; $k = 1,2,\cdots,K$; $K$ is the total number of frequency indexes; $X_r(k)$ is the amplitude of the STFT of the rth frame of the audio signal at frequency index $k$.

Figure 3 uses the distorted electric guitar and the saxophone as a comparison. The electric guitar has a clearer timbre, while the saxophone has a more turbid tone, and the spectral roll-off value of the electric guitar in the figure is significantly higher than that of the saxophone.

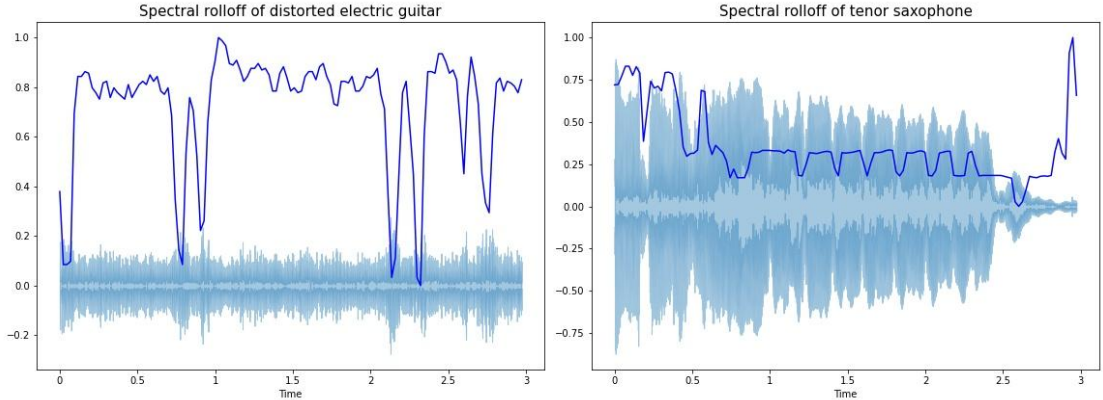

Figure 3. Spectral roll-off for distorted electric guitar and tenor saxophone.

The audio signal of music can be divided into two parts: harmonics and percussive[20]. Harmonics are sounds that have a certain pitch in the subjective perception of the human ear, such as the sound of a piano. In contrast, percussive usually originates from the sound produced by the collision of two objects, and its significant characteristic is that there is no concept of pitch, such as the sound of a drum[21]. Many sounds in the real world are a mixture of harmonics and percussive. In this paper, the music audio is decomposed into pure harmonic and percussive signals by the Har-monic-percussive Source Separation (HPSS) algorithm[22,23].

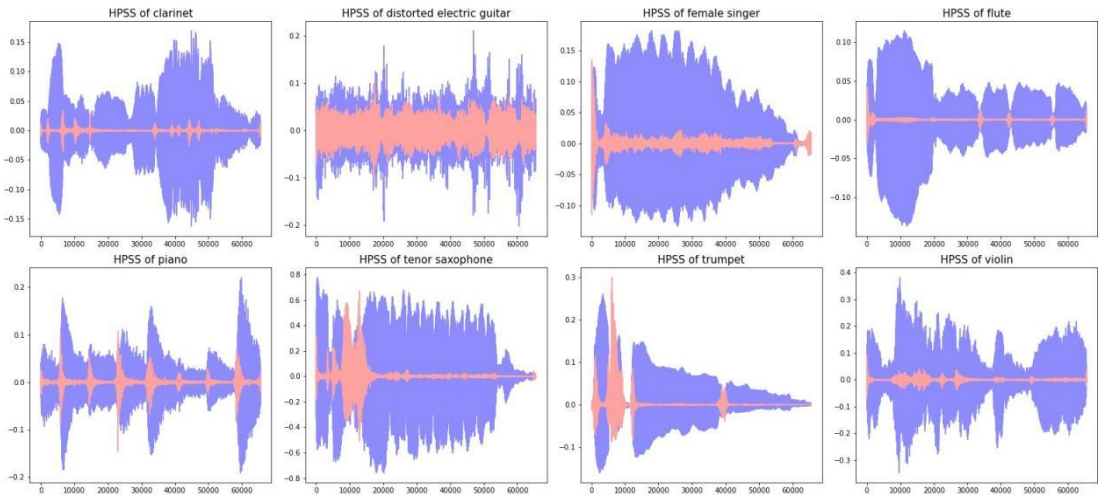

Figure 4. Harmonic and percussion signals separated by HPSS for different instruments.

*2.2. Time Domain Feature Extraction*

The time-domain features reflect the dynamic changes of the sound[24]. The different ways of coupling the excitation and resonance parts of the instrument make the envelopes of the single tone start and end of the instrument different, for example, the different resonances caused by string vibrations in different string instruments and the different nonlinear feedback caused by different overtones in wind instruments can cause differences in the envelopes.

## 2.2.1. Extraction of Zero Crossing Rate

The zero-crossing rate refers to the number of times a waveform crosses the horizontal time axis. This feature can characterize the pitch of the audio as well as the percussive sound[25]. The zero-crossing times of a frame of audio signal can be calculated by formula (2-6)[26]:

$$Z = \frac{1}{2}\sum_{n=0}^{w_l-1}|sgn[y(n)] - sgn[y(n-1)]|, \#(6)$$

where $y(n)$ is the amplitude of one frame of the audio signal, $w_l$ is the frame length, $n = 1,2,\cdots,w_l$. $sgn[\cdot]$ is a symbolic function:

$$sgn[x] = \begin{cases} 1 & x \geq 0 \\ -1 & x < 0 \end{cases}. \#(7)$$

In general, the clearer the timbre, the stronger the periodicity of the audio and the lower its zero-crossing rate[27].

## 2.2.2. Extraction of Root Mean Square (RMS) energy

The amplitude envelope is the macroscopic detection of a waveform, and a common way to calculate the amplitude envelope is the RMS algorithm[12]:

$$RMS = \sqrt{\frac{1}{L}\sum_{n=0}^{L}x^2(n)}. \#(8)$$

The RMS value is usually used to characterize the degree of loudness of audio signals perceived by the human auditory system[28].

## *2.3. Cepstral Domain Feature Extraction*

The cepstral coefficient is a way to represent resonance peak[29]. The cepstral coefficient containing the semaphore y(n) is defined as:

$$c(n) = F^{-1}\{log\,|F\{y_n\}|\}, \#(9)$$

where $F$ denotes the discrete Fourier transform.

## 2.3.1. Extraction of Mel-Frequency Cepstral Coefficients (MFCC)

MFCC analysis is based on the principle of human hearing, i.e., the spectrum of audio is analyzed based on the results of human hearing experiments in order to expect good audio characteristics[30,31].

The nonlinear characteristic of the Mayer frequency can be expressed as:

$$F_{mel} = 2595\,log\left(1 + \frac{f}{700}\right). \#(10)$$

Firstly, square the STFT spectrum to obtain the energy spectrum, and then use M equal-quality Mel bandpass filters to filter the energy spectrum to obtain the output $x'(k)$ of the $k$th filter of the filter bank. Finally, take the logarithm of $x'(k)$ to obtain the logarithmic power spectrum, and then perform inverse discrete cosine transform to obtain L MFCC coefficients. The calculation formula is:

$$C_n = \sum_{k=1}^{M} \log x'(k) \cos\left[\frac{\pi(k-0.5)n}{M}\right] \quad (n = 1,2,...,L) . \#(11)$$

In this paper, we take $L = 12$.

*2.4. Multi-channel Feature Fusion*

In order to prevent the input of the model from being too complex, we performs feature fusion on the input vector, which is simplified to a three-channel input. Since audio features are highly temporal, we use the concatenate method for fusion[32], and the fusion algorithm is given by the following equation:

$$V_T^a = v_T^1 \oplus v_T^2 \oplus \cdots \oplus v_T^a, \#(12)$$

$$V_F^b = v_F^1 \oplus v_F^2 \oplus \cdots \oplus v_F^b, \#(13)$$

$$V_C^c = v_C^1 \oplus v_C^2 \oplus \cdots \oplus v_C^c. \#(14)$$

where $V_T^a$, $V_F^a$, and $V_C^a$ are used as the input vectors of the first, second, and third channels to represent the feature fusion in the time domain, frequency domain, and cepstrum, and $a$, $b$, and $c$ represent the vector dimensions of each channel after feature fusion.

## 3. Instrument Classification Based on XGBoost

The XGBoost algorithm is an improved algorithm for the gradient boosting deci-sion tree(GBDT)[33-35], which introduces the idea of regularization to reduce the complexity of the tree and thus obtain better model performance.

*3.1. Introduction of Regular Term*

In this paper, we use the audio feature parameter vector $X$ as the input variable of a single decision tree, and the variable space of the classifier is divided into multiple regions according to the instrument type. For the input instance $x_i$, the set of decision trees with output values $\hat{y}_i$ can be predicted using K additive functions as follows.

$$\hat{y}_i = \sum_{k=1}^{K} f_k(x_i), f_k \in F, \#(15)$$

where $F$ is the CART space and each tree $f_k$ corresponds to an independent tree structure $q$ and the weights of the leaf nodes $\omega$. The classification result is obtained by classifying the given audio features to a given leaf node and calculating the total score of each leaf node by $\omega$. According to the general idea of regularized learning[35], a regularization term is added to the objective function $L$.

$$L = \sum_{i=1}^{n} l(\hat{y}_i, y_i) + \sum_{k=1}^{K} \Omega(f), \#(16)$$

where $\Omega(f_k)$ denotes the regular term on the kth tree, as follows:

$$\Omega(f_k) = \xi T + \frac{1}{2}\xi\|\omega\|^2, \#(17)$$

$l$ is the differentiable convex loss function, and the squared loss function is chosen in this paper; $\xi$ is the penalty coefficient of the regularization term, $\xi T$ and $\xi\|\omega\|^2$ are calculated for each leaf involving additional and extreme weights.

*3.2. Solving for the Optimal Tree Structure*

Since the training method of the model is additive, the instrument prediction result of the $t-1$th iteration can be adjusted according to the latest tree $f_t$ to represent the objective function of the current tth iteration. The iterative process is shown in Figure 5:

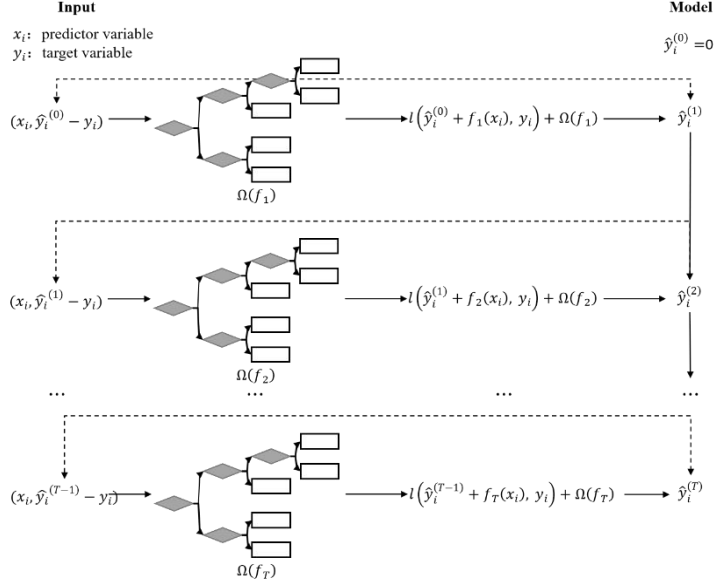

**Figure 5.** Iterative process of XGBoost algorithm.

The iteration formula is as follows:

$$L^{(t)} = \sum_{i=1}^{n} l\left(\hat{y}_i^{(t-1)} + f_t(x_i), y_i\right) + \Omega(f_t). \#(18)$$

Performing the second-order Taylor expansion on Eqs.(18), the objective function can be approximated by parameterization as

$$L^{(t)} \cong \sum_{i=1}^{n}\left[g_i f_t(x_i) + \frac{1}{2} h_i f_t^2(x_i)\right] + \Omega(f_t), \#(19)$$

where $g_i = \partial_{\hat{y}^{(t-1)}} l(y_i, \hat{y}_i^{(t-1)})$, $h_i = \partial_{\hat{y}^{(t-1)}}^2 l(y_i, \hat{y}_i^{(t-1)})$.

For a tree $f_k(x)$ with $T$ leaf nodes, it can be represented by the score vector $\omega_q(x)$ for each leaf, and then expand the second term of formula (19) to sum the leaf nodes, the original formula becomes for:

$$L^{(t)} \cong \sum_{j=1}^{T}\left[G_j \omega_i + \frac{1}{2}(H_j + \xi)\omega_j^2\right] + \xi T, \#(20)$$

where $G_j = \sum_{j \in I_j} g_i$, $H_j = \sum_{j \in I_j} h_i$, $I_j = \{i|q(x_i) = j\}$ is the instance of leaf $j$.

For the model in this paper, all decision trees adopt the same fixed structure, and the objective function is minimized as:

$$\frac{\partial L^{(t)}}{\partial \omega_j} = G_j + (H_j + \lambda)\omega_j = 0. \#(21)$$

The optimal weight of leaf $j$ is solved as:

$$\omega^* = -\frac{G_j}{H_j + \xi}. \#(22)$$

Bringing this formula into formula (20), the objective function of the optimal tree structure becomes:

$$L^{(t)} \cong -\frac{1}{2}\sum_{j=1}^{T}\frac{G_j^2}{H_j + \xi} + \xi T. \#(23)$$

Formula (3-9) is used to evaluate the effectiveness of each split node. The algorithm is performed on all possible segmentations of all features, and then approximated by a global greedy algorithm, which improves the accuracy of the classification results.

## 4. Experimental Process and Result Analysis

In this paper, we use the Medley-solos-DB dataset as an experimental sample library for instrument recognition for training and validating instrument recognition models[36,37]. The Medley-solos-DB is a cross-collection dataset consisting of 21572 audio files of 2972ms, each sampled at 44.1 kHz with 32-bit depth, and a sampling scheme conforms to the cross-collection method of Bogdanov et al.[38] and contains a total of eight timbres for clarinet, distorted electric guitar, female singer, flute, piano, tenor saxophone, trumpet and violin.

*4.1. Preprocessing*

Since each audio in the dataset is sampled with the same sampling rate and precision, the steps of audio sampling and quantization are omitted in this paper. The two steps of preprocessing are as follows:

4.1.1. Feature Normalization

Feature normalization is to improve the accuracy of the model and improve the convergence speed of the model. Since the feature vectors do not involve covariance and normal distribution, the following Min-Max Scaling algorithm is used for normalization in this paper.

4.1.2. Frame Splitting

The feature extraction of music signal is based on the steady-state signal. There-fore, before extracting the features of the music signal, it is necessary to perform frame segmentation processing to segment the signal into a small segment of the signal with stable statistical features. When calculating the STFT spectrum in this paper, the frame length is selected as 23ms. In addition, the Hamming window is used for windowing to reduce the spectral leakage error caused by not satisfying the period truncation after the frame splitting process.

The Hamming window formula is as follows:

$$\omega(n) = \begin{cases} 0.54 - 0.46 \cos\left[\dfrac{2\pi n}{N-1}\right] & 0 < n < N-1 \\ 0 & \text{else} \end{cases} . \#(4-2)$$

*4.2. Model Hyperparameters*

The Hyperparameters in the model use typical values and are manually adjusted to avoid underfitting and overfitting, and the specific hyperparameters were taken as shown in Table 1.

**Table 1.** Hyperparameter selection of XGBoost model

| Hyperparameter | Meaning | Value |
| --- | --- | --- |
| Learning_rate | Step shrinkage during iteration | 0.05 |
| n_estimators | Number of gradient boosted trees | 100 |
| max_depth | Max tree depth for base learners | 6 |
| min_child_weight | Minimum sum of instance weight needed in a child | 1 |
| subsample | Subsample ratio of the training instance | 0.8 |

*4.3. Experimental results and analysis*

In this paper, 70% of the data are randomly selected as the training set and the remaining 30% as the training set. Finally, the classification accuracy of the algorithm reached 97.65%, and the confusion matrix is shown in Figure 6.

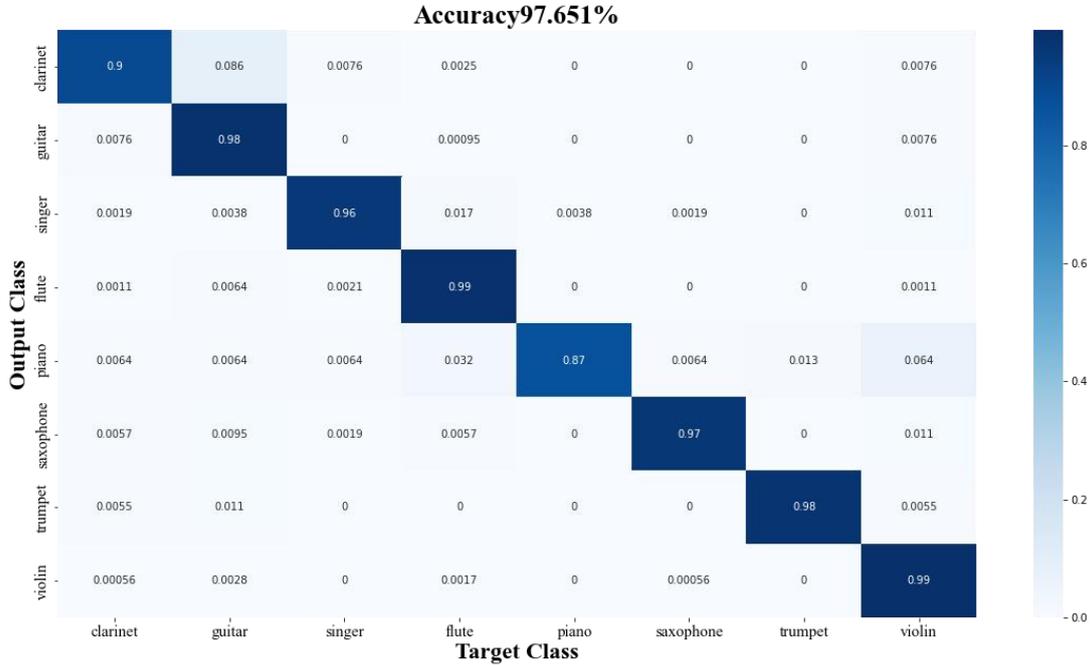

**Figure 6.** Confusion matrix of the algorithm.

In order to prove the effectiveness of the time, frequency and cepstrum domain feature selected in this paper in characterizing timbre, we tried different feature combinations for comparison. All feature combinations are classified using the XGBoost model with the same hyperparameters and loss functions. The classification results of various feature combinations are shown in Table 2.

Table 2. Classification accuracy under different feature combinations[2]

| number | Feature Domain | dimension | Accuracy (%) |
|---|---|---|---|
| A | Time | 4 | 67.10 |
| B | Frequency | 8 | 87.78 |
| C | Cepstrum | 24 | 95.23 |
| D | Time, Frequency | 12 | 89.81 |
| E | Time, Cepstrum | 28 | 96.20 |
| F | Frequency, Cepstrum | 32 | 96.66 |
| G | Time, Frequency, Cepstrum [selected combination] | 36 | 97.65 |
| H | Time, Frequency, Cepstrum, Autocorrelation feature | 40 | 96.68 |

From the comparison, it can be seen that among the 8 feature combinations, the combination G including the time domain, frequency domain and cepstral domain achieves the best accuracy rate of 97.65%.

However, this optimal combination may just result from the increasing number of features. In order to falsify this hypothesis, we also designed a special combination H with 40-dimensional features, which adds two autocorrelated features characterizing timbre, beat rate and intrinsic strength[10]. However, although this set of data has the largest number of features, the accuracy rate is not as good as that of 36-dimensional feature combinations. Therefore, it can be concluded that the combination of audio features selected in this paper has superior performance in musical instrument classification.

In addition, we evaluate the performance of a variety of commonly used machine learning models, including Naive Bayes, Stochastic Gradient Descent(SGD), KNN, De-cision Trees, Support Vector Machine(SVM) and Logistic Regression(LR).

Table 3. Classification accuracy of different classification models

| Model | Accuracy(%) |
|---|---|
| Naive Bayes | 67.29 |
| Stochastic Gradient Descent | 40.22 |
| KNN | 64.14 |
| Decision trees | 91.73 |
| SVM | 56.83 |
| Logistic Regression | 53.93 |
| XGBoost [selected model] | 97.65 |

The results show that the classification accuracy of XGBoost model is 97.65%, which is better than other classification models. Compared with the results of other studies on the same dataset, the accuracy of this model is improved by about 22.65% compared with the pitch spiral convolutional network model [36]; compared with the joint time-frequency scattering model [37], the accuracy is improved About 19.65%.

---

[2] The code used to calculate Table 2, 3 is available at the URL:
https://github.com/Jay-Codeman/Musical-Instrument-Recognition-by-XGBoost.

## 5. Discussion

The expressiveness of the features and the performance of the classifier are the two most significant factors that affect the results of musical instrument classification. In this paper, 18 timbre-related features in the time, frequency and cepstrum domain are fused in multiple channels, and through comparative experiments, we verify that this feature combination has the best performance in musical instrument classification. We also verify that the improvement in the accuracy of the algorithm is not just due to the increase in the number of feature dimensions. The accuracy of the algorithm in this paper on the public dataset medley-solos-DB reached 97. 65%.

We also compare multiple machine learning classification models and verify that the XGBoost model has the highest accuracy. By comparing with the results of other studies on the same dataset, we verify the performance superiority of our model on musical instrument classification.

## 6. Conclusions

Our research provides a reference for the selection of classifiers for musical instrument recognition, and also provides ideas for feature selection and fusion in audio feature engineering. In future research, the model needs to be further optimized in terms of hyperparameters, error and bias analysis according to the needs of real application scenarios.

**Author Contributions:** Conceptualization, Y.L. and W.C.; methodology, Y.L. and Y.Y.; software, Y.L.; validation, Y.L., Y.Y. and W.C.; formal analysis, W.C.; investigation, Y.L. and W.C.; resources, Q.Z.; data curation, Y.L.; writing—original draft preparation, Y.L.; writing—review and editing, Y.L. , Y.Y. and Q.Z.; visualization, W.C.; supervision, Y.Y. and Q.Z.; project administration, Y.L. and Q.Z.; funding acquisition, Y.L. and W.C. All authors have read and agreed to the published version of the manuscript.

**Institutional Review Board Statement:** Not applicable.

**Informed Consent Statement:** Not applicable.

The datasets generated during analysed during the current study are available in the Medley-solos-DB repository at [10.5281/zenodo.3464194], reference number [36, 37].

**Acknowledgments:** The authors would like to thank Yanfang Yin and Qigang Zhu for the writing advice. and Yijie Liu wants to thank Li Ma, Yuan Liu and Luping Zhao for the constant support.

**Conflicts of Interest:** The Authors declare that there is no conflict of interest.

**Funding:** The authors disclosed receipt of the following financial support for the research, authorship, and publication of this article: This work was supported by the Shandong University of Science and Technology Excellent Teaching Team Support Program Sino-foreign Cooperation based on the project-driven communication engineering education teaching team[grant number JXTD20180510].

## References

1. Huo, Y. Music Personalized Label Clustering and Recommendation Visualization. *Complexity* **2021**, *2021*, 5513355, doi:10.1155/2021/5513355.
2. Wang, M.; Xiao, Y.; Zheng, W.; Xu, J.; Hsu, C.H. Tag-Based Personalized Music Recommendation. In *Proceedings of the International Symposium on Pervasive Systems, Algorithms and Networks*, 2018.
3. Datta, A.K.; Solanki, S.S.; Sengupta, R.; Chakraborty, S.; Mahto, K.; Patranabis, A.J.S.S. *Automatic Musical Instrument Recognition.* **2017**.
4. Özbek, M.E.; Systems, N.Ö.A.S.J.J.o.I.I. *Wavelet ridges for musical instrument classification.* **2012**.
5. Zlatintsi, A.; Maragos, P.J.I.T.o.A.S.; Processing, L. Multiscale Fractal Analysis of Musical Instrument Signals With Application to Recognition. **2013**, *21*, 737-748.
6. Rui, R.; Bao, C.C.J.R. Musical Instrument Classification Based on Nonlinear Recurrence Analysis and Supervised Learning. **2013**, *22*, 60-67.
7. Kostek, B.; Wieczorkowska, A.J.A.o.A. Parametric representation of musical sounds. **1997**, *22*, p.3-26.
8. Sikora, T. The MPEG-7 visual standard for content description-an overview. *IEEE Transactions on Circuits and Systems for Video Technology* **2001**, *11*, 696-702, doi:10.1109/76.927422.
9. Wieczorkowska, A.A.; Jakub. Application of temporal feature extraction to musical instrument sound recognition.
10. Brown, J.C.; Houix, O.; McAdams, S.J.T.J.o.t.A.S.o.A. Feature dependence in the automatic identification of musical woodwind instruments. **2001**, *109*, 1064-1072.
11. Martin, K.D. Sound-source recognition: A theory and computational model. Massachusetts Institute of Technology, 1999.
12. Chen, G.; Zhang, X. A voice/music classification method based on grey relational analysis. *Acoustic Technology* **2007**, 262-267.
13. VERMA, Y. A Tutorial on Spectral Feature Extraction for Audio Analytics. Available online: https://analyticsindiamag.com/a-tutorial-on-spectral-feature-extraction-for-audio-analytics/ (accessed on SEPTEMBER 19, 2021).
14. Schubert, E.; Wolfe, J.J.A.a.u.w.a. Does timbral brightness scale with frequency and spectral centroid? **2006**, *92*, 820-825.
15. Wei, L. Research on Audio Signal Classification Algorithm. Dalian University of Technology, 2009.
16. Klapuri, A.; Davy, M. Signal processing methods for music transcription. **2007**.
17. Center for Computer Research in Music and Acoustics, S.U. CCRMA SUMMER WORKSHOPS 2021. Available online: https://ccrma.stanford.edu/ (accessed on Feburay 19, 2022).
18. Kos, M.; Kai, Z.; Vlaj, D.J.D.S.P. Acoustic classification and segmentation using modified spectral

*roll-off and variance-based features.* **2013**, *23*, 659-674.

19. *Scheirer, E.; Slaney, M. Construction and evaluation of a robust multifeature speech/music discriminator. In Proceedings of the 1997 IEEE International Conference on Acoustics, Speech, and Signal Processing, 21-24 April 1997, 1997; pp. 1331-1334 vol.1332.*

20. *Lakatos, S.J.P.; Psychophysics. A common perceptual space for harmonic and percussive timbres.* **2000**, *62*, 1426-1439.

21. *Driedger, J.; Müller, M.; Ewert, S.J.I.S.P.L. Improving time-scale modification of music signals using harmonic-percussive separation.* **2013**, *21*, 105-109.

22. *Fitzgerald, D. Harmonic/percussive separation using median filtering. In Proceedings of the Proceedings of the International Conference on Digital Audio Effects (DAFx), 2010.*

23. *Driedger, J.; Müller, M.; Disch, S. Extending Harmonic-Percussive Separation of Audio Signals. In Proceedings of the ISMIR, 2014; pp. 611-616.*

24. *Joder, C.; Essid, S.; Richard, G. Temporal Integration for Audio Classification With Application to Musical Instrument Classification. IEEE Transactions on Audio, Speech, and Language Processing* **2009**, *17*, 174-186, doi:10.1109/TASL.2008.2007613.

25. *Chauhan, N.S. Audio Data Analysis Using Deep Learning with Python (Part 1). Available online: https://www.kdnuggets.com/2020/02/audio-data-analysis-deep-learning-python-part-1.html (accessed on February 19, 2022).*

26. *Sun, H.; Long, H.; Shao, Y.; Du, Q. Speech and Music Classification Algorithm Based on Zero-Crossing Rate and Spectrum. Journal of Yunnan University(Natural Sciences Edition)* **2019**, *41*, 925-931.

27. *Zhang, C.; Yang, Y.; Hu, R. Research on Audio Content Segmentation and Clustering. Computer Engineering* **2002**, 173-174.

28. *Panagiotakis, C.; Tziritas, G.J.I. A speech/music discriminator using RMS and zero-crossings.* **2002**.

29. *Yang, J. Research on the extraction method of musical instrument timbre features based on harmonic structure. Master, Harbin Institute of Technology, 2018.*

30. *Vergin, R.; O'Shaughnessy, D.; Farhat, A.J.I.T.o.s.; processing, a. Generalized mel frequency cepstral coefficients for large-vocabulary speaker-independent continuous-speech recognition.* **1999**, *7*, 525-532.

31. *Gulhane, S.R.; Badhe, S.; Shirbahadurkar, S.D. Cepstral (MFCC) Feature and Spectral (Timbral) Features Analysis for Musical Instrument Sounds. In Proceedings of the 2018 IEEE Global Conference on Wireless Computing and Networking (GCWCN), 2018.*

32. *Feature Fusion: Pointwise Addition Or Concatenate Vectors? – Deep Learning Tutorial. Available online: https://www.tutorialexample.com/feature-fusion-pointwise-addition-or-concatenate-vectors-deep-learning-tutorial/ (accessed on Apr 15, 2022).*

33. *Chapelle, O.; Chang, Y. Yahoo! learning to rank challenge overview. In Proceedings of the Proceedings of the learning to rank challenge, 2011; pp. 1-24.*

34. *Zhang, C.; Liu, C.; Zhang, X.; Almpanidis, G.J.E.S.w.A. An up-to-date comparison of state-of-the-art classification algorithms.* **2017**, *82*, 128-150.

35. *Chen, T.; Guestrin, C. Xgboost: A scalable tree boosting system. In Proceedings of the Proceedings of the 22nd acm sigkdd international conference on knowledge discovery and data mining, 2016; pp. 785-794.*

36. *Lostanlen, V.; Cella, C.-E.J.a.p.a. Deep convolutional networks on the pitch spiral for musical instrument recognition.* **2016**.

37. *Andén, J.; Lostanlen, V.; Mallat, S.J.I.T.o.S.P. Joint time–frequency scattering.* **2019**, *67*, 3704-3718.

38. *Bogdanov, D.; Porter, A.; Boyer, H.; Serra, X. Cross-collection evaluation for music classification tasks. In Proceedings of the Devaney J, Mandel MI, Turnbull D, Tzanetakis G, editors. ISMIR 2016. Proceedings of the 17th International Society for Music Information Retrieval Conference; 2016 Aug 7-11; New York City (NY).[Canada]: ISMIR; 2016. p. 379-85., 2016.*